\documentstyle[twocolumn,floats,epsfig,prl,aps]{revtex}
\begin{document}
\narrowtext
{\bf Eggert, Gustafsson, and Rommer Reply to the ``Comment on
`Phase diagram of an impurity in the spin-1/2 chain: two channel Kondo effect versus Curie 
law' '' \\ }
\author{Sebastian Eggert, David P.~Gustafsson, and Stefan Rommer}
\address{Institute of Theoretical Physics,
Chalmers University of Technology and G\"oteborg University,
S-412 96 G\"oteborg, Sweden}
\date{\today}
The Comment\cite{comment} introduces the particular point 
at $J_1 \to \infty$ as a new fixed point in the phase diagram of the 
$J_1$-$J_2$-impurity model which was analyzed in our recent Letter\cite{letter}. 
The point at $J_1 \to \infty$ is certainly worth a separate discussion which
we will give here, but our analysis comes to rather different conclusions
than Zvyagin\cite{comment}.

At $J_1\to \infty$ the three spins $\vec S_0$, $\vec S_1$, and $\vec S_N$
are strongly coupled and form a complex of total spin $s=1/2$, which 
is characterized by a {\it triplet} between the spins 
$\vec S_1$ and $\vec S_N$, which in turn is antiferromagnetically correlated
with $\vec S_0$. 
This three-spin complex is therefore {\it not} decoupled from the 
rest of the chain, but the effective coupling is given by
$2J/3$ instead. 
The point at $J_1\to \infty$ therefore does {\it not}
represent a fixed point at all since it does not correspond to a simple
boundary condition in the spin chain. 
This is in contrast to the fixed point 
${\rm O_{N-2} \otimes \case{1}{2}}$ at $J_2 \to \infty$ where the 
three-spin complex is indeed decoupled from the chain
since the spins $\vec S_1$ and $\vec S_N$
form a {\it singlet}.  The effective coupling there is given by $-J^2 J_1/4J_2^2$ 
from a {\it third} order perturbation expansion, which is always irrelevant because
this coupling is suppressed by one additional power of the cutoff due to 
the virtual excitations.  

For simplicity we will label the point at $J_1\to \infty$ by
$\rm P_{N-1}$ because three spins are effectively replaced by
one spin-1/2 complex which remains coupled to the chain.
This point $\rm P_{N-1}$ is again characterized by a logarithmically
diverging impurity susceptibility and a ferromagnetic correlation 
$\langle \vec{S}_1 \cdot \vec{S}_N \rangle > 0$, just like the fixed
point $\rm P_{N+1}$.  We therefore do not expect any phase transitions
or any discontinuities in the order parameter between the two points.

As far as the field theory description near the fixed point $\rm P_{N+1}$
is concerned, we must emphasize again that the only leading irrelevant operator
is given by $\partial_x  {\rm tr g}$\cite{eggert}.
The operators mentioned in the Comment\cite{comment} are not present 
at $\rm P_{N+1}$, since the impurity spin 
$\vec S_0$ has been absorbed in the chain
and $\vec S_0$ can therefore not be used as an independent degree of freedom 
to construct operators.

\begin{figure}
\begin{center}
\mbox{\epsfig{file=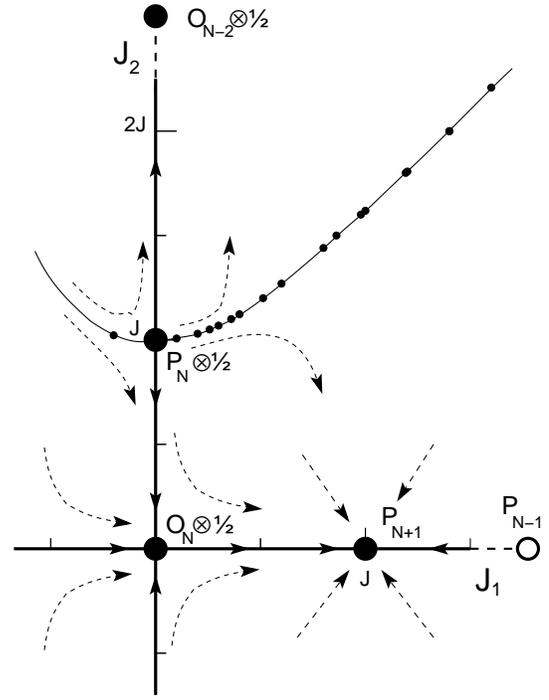,width=2.8in,angle=0}}
\end{center}
\caption{The full phase diagram of the $J_1$-$J_2$-impurity model.
Black dots are fixed points.
The point $\rm P_{N-1}$ is indicated by a circle and  
does not represent a fixed point of the model.}
\label{phasediagram}
\end{figure}

In conclusion we find that the point $\rm P_{N-1}$ is not a fixed point
and it appears to be in the same phase as $\rm P_{N+1}$.  The phase diagram
as shown in Fig.~\ref{phasediagram} is therefore complete and correct.

The comparison to the integrable impurity model\cite{frahm} made in the 
Comment\cite{comment} is certainly interesting.  However, obviously 
the integrable impurity model lives in a different parameter space, 
so that a direct comparison of the impurity susceptibilities may not be
very meaningful.  An impurity in a free Fermion model (xx-model) also 
gives only limited insight since the scaling 
dimensions are fundamentally different in interacting systems.
 However, it would be interesting to examine the integrable model\cite{frahm}
in more detail with field theory methods in an expanded parameter space, 
which, however, is not as trivial as indicated in the Comment\cite{comment}.
This would answer the question if it may belong into one of the phases that were
discussed in our Letter\cite{letter} or if it may correspond to a
non-generic unstable
multi-critical point as has been found for a related integrable
impurity model\cite{sorensen}.\\~\\
{Sebastian Eggert, David P.~Gustafsson, and Stefan Rommer}


\end{document}